\documentclass[12pt,a4paper]{article}
\usepackage{graphics,amsfonts}

\newcommand{\be}{\begin{equation}}
\newcommand{\ee}{\end{equation}}
\newcommand{\eel}[1]{\label{#1}\end{equation}}
\newcommand{\bea}{\begin{eqnarray}}
\newcommand{\eea}{\end{eqnarray}}
\newcommand{\eeal}[1]{\label{#1}\end{eqnarray}}
\newcommand{\baq}{\begin{equation}\begin{array}{rcl}}
\newcommand{\eaq}{\end{array}\end{equation}}
\newcommand{\eaql}[1]{\end{array}\label{#1}\end{equation}}
\newcommand{\beac}{\begin{equation}\begin{array}{rcl}}
\newcommand{\eeacn}[1]{\end{array}\label{#1}\end{equation}}
\newcommand{\ba}{\begin{array}}
\newcommand{\ea}{\end{array}}

\newcommand{\beq}{\begin{eqnarray}}
\newcommand{\eeq}{\end{eqnarray}}
\newcommand{\beqno}{\begin{eqnarray*}}
\newcommand{\eeqno}{\end{eqnarray*}}

\newcommand{\pa}{\partial}

\newcommand{\preprint}[1]{\begin{table}[t]  
           \begin{flushright}               
           \begin{large}{#1}\end{large}     
           \end{flushright}                 
           \end{table}}                     

\def\xop{\hat{x}}
\def\Aop{\hat{A}}
\def\Uop{\hat{U}}
\def\Vop{\hat{V}}

\def\half{{1\over 2}}

\input{psfig}
\begin{document}
\begin{titlepage}
{\small{
\preprint{hep-th/0105089\\TAUP-2675-2001}
}}
\vspace{2cm}

\begin{center}

{\Large \bf UV/IR connection, a matrix perspective}\footnote{
The work of Y.K and J.S is supported in part by the US-Israel
Binational Science Foundation,
by GIF - the German-Israeli Foundation for Scientific Research,
and by the Israel Science Foundation.
}\\

\vspace{1.5cm}

\bf{Y. Kinar${}^{1}$, G. Lifschytz${}^{2}$ and J. Sonnenschein${}^1$}

\vspace{4mm}

${}^1${\small \sl Raymond and Beverly Sackler Faculty of Exact
  Sciences\\
  School of Physics and Astronomy\\
  Tel-Aviv University, Ramat Aviv, 69978 Israel \\
  e-mail: \{yaronki,cobi\}\verb+@+post.tau.ac.il}
\vspace{6mm}

${}^2${\small \sl Department of Mathematics and Physics \\
  University of Haifa at Oranim, Qiryat-Tivon, Israel\\
  e-mail: giladl@research.haifa.ac.il}

\end{center}
\vskip 0.5 cm
\begin{abstract}
We show that the matrix formulation of non-commutative field theories 
is equivalent, in the continuum, to a formulation in a mixed 
configuration-momentum space. In this formulation, the non-locality
of the interactions, that leads  to the IR/UV mixing, becomes
transparent. We  clarify the relation between long range effects (and
IR divergences) and the non-planarity of the corresponding Feynman diagrams.

\end{abstract}
\end{titlepage}

\section{Introduction}
While much work has been devoted to the understanding of field
theories on non-commutative spaces (see \cite{Conn}), in
particular after those theories were related to string theory
(\cite{CoDoSc},\cite{dh},\cite{ShJab},\cite{BFSS},\cite{SeWi},\cite{sei}), some issues remain unclear. An
important example of such an issue is the so called  
"IR/UV mixing" \cite{MiRaSe}. This peculiar behaviour of field theories 
on non-commutative spaces is manifested through the existence of a class
of non-planar diagrams which are divergent for vanishing external
momenta. The divergence originates from the phase of these non-planar
diagrams that depends on both the external  and internal momenta.
The integration of the latter in loops results in UV divergences
which are the source of the ``IR divergences" of  the non-planar diagrams. 
 
The UV/IR mixing was further discussed both for scalar field theories
(\cite{RaSe}-\cite{GoLaLo}) and for gauge theories which can be
associated with the open string perspective
(\cite{MaSuTo}-\cite{KiLePa}). The immediate question of
the effect of this mixing on the renormalizability of these theories
was also intensively considered (\cite{CheRo1}-\cite{KhTr}). 

In this note we would like to further investigate the origin 
of the IR divergences and clarify their relations to
non-commutativity, non-locality of the interaction, and non-planarity
of the relevant diagrams.  

We use the known mapping between a non-commutative field theory and a
 matrix theory. We make the observation
 that the formulation in terms of  $N\times N$ matrices corresponds,
 in the continuum picture, to expressing the fields as functions of
 coordinates of a mixed configuration-momentum space. By using
 matrices of finite size we regularize both the UV and the IR divergences, 
since the maximal value of the momentum and the area are proportional
to $\sqrt{N}$ and $N$ respectively. 
We show, in the mixed configuration-momentum space, that the interactions of 
scalar field theories defined on non-commutative spaces are  non-local. 
 The non-locality renders the interactions into long range ones even for massive fields. 
For space with infinite volume,  these long range interactions are behind
 the IR divergences associated with certain 
non-planar diagrams. We show  that the effect of the 
non local interactions is  to bring the interaction points closer 
to each other. 
This behaviour is the source of the  ``mixing''  of the IR divergences with the contributions
of high momentum that leads usually  to  UV divergences.
We explain why this effect exists (and is natural) only
for certain non-planar diagrams in the matrix theory.

In terms of the dependence on the cutoff $N$, we show that the UV
divergent planar diagrams have the same $N$ dependence as IR divergent
non-planar diagrams of the same loop order. 
We demonstrate this behaviour for the case of the two point function.
Each planar diagram in this case 
 has a higher $N$ dependence than that of a corresponding non-planar one
due to the summation of all possible indices  in closed internal
loops. However, this factor is compensated by the multiplicity of
non-planar diagrams. Thus, even in the $\theta \rightarrow \infty$ limit some
non-planar diagrams contribute.  

The  paper is organized as follows. In section 2 we describe the 
basic setup, namely, the operator formulation of the 
scalar field theory on non-commutative geometry. Anticipating 
the structure of the matrix description,  we rewrite in section 3 
the action in a mixed configuration-momentum space representation.
In section 4 we determine the propagator in the operator (matrix) framework,
and show, using the 'shift' and 'clock' matrices, that it indeed corresponds
to the mixed description.
We state in section 5 the Feynman  rules in the latter description as well as
in the matrix formulation. 
 Section 6 is devoted to the the origin of the IR divergences.
In particular, it is shown  that the noncommutativity 
introduces
long range interactions, and explained why the IR divergences are associated with 
non-planar diagrams. We determine the dependence of the various diagrams on $N$.
An example is described in the appendix. 
It includes the leading order corrections to the $\phi^3$ two point
function from both planar and non-planar diagram and determines the corresponding phases.  

\section{The basic setup}

We start with a brief description of the basic setup of the generic
theory we study, noting our conventions in passing. Consider the
non-commutative  $\phi^K$ field theory on a
(d+1)-dimensional plane , whose Euclidean action is
\be
S = \int dt d^dx \left(\half (\pa_\mu\phi)^2 + \half M^2 \phi^2 +
  \lambda \phi^K \right)
\ee
where the products are Moyal $*$-products
\be
\label{Moyal}
(\phi_1 * \phi_2)(x) = \exp{({i \over 2} \theta^{\mu \nu} \pa_\mu^y
  \pa_\nu^z)} \phi_1(y) \phi_2(z) | _{y=z=x}
\ee
In particular
\be
e^{ikx} * e^{ipx} = \exp{(-{i \over 2} k \times p)} e^{i(p+k)x} 
\ee
or 
\be
e^{ikx} * e^{ipx} = \exp{(-i k \times p)} e^{ipx} * e^{ikx}
\ee
where
\be
k \times p \equiv k_\mu \theta^{\mu \nu} p_\nu
\ee
We  Fourier transform $\phi(x)$
\be
\tilde{\phi}(k) = {1 \over \sqrt{(2 \pi)^d}} \int d^dx e^{-ikx} \phi(x)
\ee
and represent the Moyal algebra by the operators $\Aop(\xop)$
defined as (see for instance \cite{gms},\cite{IsKaKi},\cite{GrNe})
\be
\label{oper} 
\Aop(\xop) =  {1 \over \sqrt{(2 \pi)^d}} \int d^dk e^{ik \xop}
  \tilde{\phi}(k)
\ee
where $\xop^\mu$ are operators satisfying the commutation relations
\be
[\xop^\mu,\xop^\nu] = - i \theta^{\mu\nu}
\ee

We  now formulate the non-commutative field theory using the
operators formalism, according to the mapping
\beq
\phi_1 * \phi_2 & \rightarrow & \Aop \cdot \hat{B} \nonumber \\
{1 \over 2 \pi \mbox{Pf}(\theta)} \int d^dx & \rightarrow & Tr
\nonumber \\
i \theta^{\mu \nu} \pa_\nu & \rightarrow & [\xop^\mu,\cdot]
\eeq

For simplicity, we assume there are only two non-commutative coordinates,
$rank(\theta)=2$, $[\xop^1,\xop^2]= - i \theta$, and ignore all the other
 (commutative) coordinates. We define the exponents
\be
\Uop \equiv \exp{(-i \xop^1)} ;\qquad  \Vop \equiv \exp{(-i \xop^2)}
\ee
satisfying
\be 
\Uop \Vop  = e^{-i \theta} \Vop \Uop
\ee
Using these definitions   the Fourier transform takes the following
form 
\beq
{\tilde{\Aop}}(p,q) &= &Tr \left(e^{-i (p \xop^1 + q \xop^2)} \Aop
\right)\nonumber \\
& = & Tr \left( \exp{({i \over 2} \theta pq)} e^{-i p \xop^1} e
  ^{-i q \xop^2} \Aop \right) \nonumber \\
& \equiv & Tr \left( :\Uop^p \Vop^q: \Aop \right)
\eeq
and the action 
\be
\label{action}
\int dt Tr \left( \theta^{-2} [\xop^\mu,\Aop]^2 + \half M^2 \Aop^2
    +\lambda \Aop^K \right)
\ee

\section{Mixed configuration-momentum space}

The interesting relation between momentum and
size in non-commutative field theories (\cite{BiSu}),
guides us to 
consider the mixed configuration--momentum space. Working in two
space dimensions $x$ and $y$, one can Fourier transform the action
only in the $x$-direction. The kinetic term becomes
\be
\label{mixkin}
S_{kin} = \half \int dp_x dy \phi(\tilde{p_x},y) \left(p_x^2 - \partial_y^2
\right)  \phi(\tilde{p_x},y)
\ee
The notation is such that from now on, we use $\phi$ for the field in all 
representations, and
indicate momentum-space coordinates by a tilde - $\tilde{p}$ and 
configuration space coordinate
by a non-tilde variable - $y$. The
mass term remains the same while
the interaction becomes (considering $\phi^3$ as a simple example) - 

\beqno
S_{int} &=& \int dx dy \phi(x,y) * \phi(x,y) * \phi(x,y) \\
        &\sim& \int dx dy dp_1 dp_2 dp_3  (e^{i p_1 x}
        \phi(\tilde{p_1},y)) * (e^{i p_2 x} \phi(\tilde{p_2},y)) * (e^{i
          p_3 x} \phi(\tilde{p_3},y)) \\
        &=& \int dy dp_1 dp_2 \phi(\tilde{p_1},y-\theta p_1/2)
        \phi(\tilde{p_2},y-\theta p_1- \theta p_2/2)
        \phi(-\tilde{p_1}-\tilde{p_2},y-\theta p_1/2- \theta p_2/2)
\eeqno
We can further change our notations and write
\beq
A(z,w) &\equiv& \phi(\frac{1}{\theta}(\tilde{z}-\tilde{w}),\half(z+w)) \\
\phi(\tilde{p},x) &=& A(x+{{\theta p} \over 2},x - {{\theta p} \over 2})
\eeq
and the interaction term becomes
\beqno
S_{int} &=& \int dy dp_1 dp_2 A(y,y- \theta p_1) A(y- \theta p_1,y-
\theta p_1-\theta p_2) A(y-\theta p_1-\theta p_2,y) \\
        &=& Tr A^3
\eeqno
where the last line can be considered formally or via
discretization. From now on we will put $\theta=1$ in our
expressions (unless stated otherwise), 
it can be easily inserted back to restore the
dimensionality.  The last expression clearly shows that the action in
the mixed configuration-momentum space has a matrix structure. In the
next section we will show that the familiar 'clock' and 'shift'
representation of the operators $\Uop$ and $\Vop$ indeed leads to this
mixed space. We will further discuss the non-locality of the
interaction term in section \ref{sec:divs}. 

\section{Operator (matrix) space}

In order to consider the perturbative behaviour of the theory in the 
 operator representation, the first step we take is 
to regularize any divergence by considering a finite dimensional
Hilbert space. In other words,
the various operators are now $N \times N$ matrices. 
Eventually, we will let $N$ go to infinity. The degrees of freedom are the elements of
the matrix $\Aop$.
 
It is well known that there is no finite dimensional matrix
representation for the non-commuting operators $\xop^1$ and
$\xop^2$. However, any $N \times N$ matrix can be expanded in a finite
power series in two unitary matrices which satisfy
\be
UV = e^{-2 \pi i / N}VU = \omega VU
\ee
and are therefore related to $\Vop$ and $\Uop$ via (restoring $\theta$
for a moment)
\beq
\Uop &=& U^{\sqrt{\frac{ \theta N}{2 \pi}}} \nonumber  \\
\Vop &=& V^{\sqrt{\frac{\theta  N}{2 \pi}}}
\eeq

The propagators are
\be
\label{corr}
<A_{ij} A_{kl}> = \frac{1}{N}\sum_{m n} :U^m V^n:_{ij} :U^{-m}
V^{-n}:_{kl} \frac{1}{(m^2 + n^2)/N + M^2}
\ee

$V$ and $U$ can be represented by the $N
\times N$  'clock' and 'shift' matrices (this was also used in 
\cite{LaLiSz},\cite{KrRaSh}):

\beq
U_{i,i+1} = 1 && i = 1 ... N-1\nonumber \\
U_{N,1}   = 1 &&\nonumber \\
V_{i,i}   = \omega^{i-1}  &&\omega=e^{-2 \pi i / N}
\eeq

Using these matrices, we can now go back
to the definition of the operator representation of our algebra,
(\ref{oper}), and rewrite it as 

\beq
A_{ij} &=& {1 \over {2 \pi N}} \sum_{mn} :U^m V^n:_{ij}
\phi(\tilde{m},\tilde{n}) \nonumber \\
        &=& {1 \over {2 \pi N}} \sum_n \omega^{\half n(i+j-2)}
          \phi(\tilde{j}-\tilde{i},\tilde{n})=
          \frac{1}{\sqrt{2\pi N}}\phi(\tilde{j}-\tilde{i},(j+i)/2) 
\label{aphi}
\eeq
which shows, as promised, that the matrix indices in the operator
formalism correspond to the
mixed configuration-momentum space. If we change the
indexing of the matrix element $(i,j)$ to $(j-i \mbox{ mod } N,(i+j)/2
\mbox{ mod } N)$, the first index corresponds to momentum space while
 the second to configuration space.

Inserting $U$ and $V$ into the propagators (\ref{corr}) gives

\be
<A_{ij}A_{kl}> = \delta_{i-j,l-k} \frac{1}{N}
                \sum_n\frac{w^{n(j-k)}}{M^2 + ((j-i)^2 + n^2)/N}
\ee
Taking the limit $N \rightarrow \infty$ we should also scale $i,j,k$ and
$l$. Explicitly, we take the combinations $x \equiv 2 \pi (j-k)/\sqrt{N}$
and $ p \equiv (j-i)/\sqrt{N}$ fixed while taking the limit and write
$q \equiv (k-l)/\sqrt{N}$ (which must be also fixed). The sum in
the propagator becomes (this is just the usual propagator)
\be
\label{fullphi}
 \frac{1}{{N}} \delta(p+q) \int_{-\infty}^\infty dk
 \frac{e^{ikx}}{p^2 +M^2 + k^2} = \frac{\delta(p+q)}{N}
\frac{e^{-(p^2 +M^2)|x|}}{p^2 +M^2}
\ee
Note that an extra factor of $\sqrt{N}$ 
comes from the $\delta$-function. This expression for the propagator
is, of course, the one we get for the mixed space up to the power of 
$\frac{1}{N} $ that comes from  equation (\ref{aphi}). 

One can easily see that the operator formalism, using the 'clock' and
'shift' matrices, corresponds to the mixed-space formalism 
 by considering the
kinetic term - $[\xop^\mu,\Aop]^2$.
This term comprises of a diagonal term
from $\xop^2 = i\mbox{log} V$ corresponding to the momentum coordinate
and a non-diagonal term from $\xop^1 = i\mbox{log} U$ corresponding to
the configuration coordinate. This is the same as in the mixed-space
representation, as equation (\ref{mixkin}) shows.

\section{Feynman diagrams}
\label{sec:feyn}

If one works with the Moyal product representation of the
algebra, that is with functions on ${\mathbb R}^d$ and the multiplication rule
(\ref{Moyal}), it is possible to define momentum-space Feynman rules,
which differ from the commutative ones only in the K-point vertex,
which acquires a phase (depending on the {\em{order}} of the momenta
flowing into the vertex) : 
\be  
V(k_1,k_2 ...,k_K) = \lambda e^{-{i \over 2} \sum_{i<j} k_i \times
  k_j} \delta(\sum k_i)
\ee

Considering various Feynman graphs, it is shown in \cite{MiRaSe} that
in planar  graphs these phases add up so that the overall phase depends only
on the external momenta, while in non-planar graphs, one is left with 
phases depending on internal (integrated) momenta. These phases may
give rise to IR divergences in non-planar graphs (The UV/IR mixing).

When one works in a mixed configuration space
the propagator takes the form (see equation (\ref{fullphi}))
\begin{equation}
<\phi(\tilde{p},x)\phi(\tilde{q},y)>=\delta(p+q)
\frac{e^{-(p^2 +M^2)|x-y|}}{p^2 +M^2}
\label{propcont}
\end{equation}
The interaction vertex (for $\phi^{n}$) takes the form

\beq
\int dx d^{n}p \;\delta(\sum_{i=1}^{n}p_{i})\phi(\tilde{p}_{1},x)\phi(\tilde{p}_{2},x-(p_1 +p_2))\phi(\tilde{p}_{3},x-(p_1 +2p_{2}+p_{3}))\cdots\nonumber\\
\phi(\tilde{p}_{j},x-(p_1 +2\sum_{k=2}^{j-1} p_{k}+p_{j}))
\cdots \phi(\tilde{p}_{n}, x-(\sum_{k=2}^{n-1}p_{k}))
\label{vertcont}
\eeq

Given a (non) planar diagram in the momentum space representation of a
non-commutative theory, it clearly remains such a diagram  
 in the mixed configuration-momentum space. Thus, the (non) planar diagrams of 
the momentum spaces representations are the (non) planar diagram of the matrix
formalism. Considering the Feynman rules for the matrix formalism, it
is clear that the phases in the non-planar diagrams (before performing
any sums on external momenta) may come only from the propagators
(\ref{corr}) as the vertices clearly do not carry any phase. For future use,
we wish to consider the simpler case where there is no kinetic term in
the action. This may be considered as the $\theta \rightarrow \infty$
limit of the action  (\ref{action})\footnote{Note that for generic
  values of the external momenta this limit is dominated by planar
  diagrams. However, the IR divergences we are interested in, appear
  in special values of the external momenta}, which reduces to 

\begin{equation}
S=\int dt (M^2 Tr A^2 + \lambda Tr A^K)
\end{equation}

The free (tree level) propagators for this theory are simply

\be
\label{propA}
<A_{ij} A_{kl}> = \frac{1}{N}\sum_{m n} :U^m V^n:_{ij} :U^{-m}
V^{-n}:_{kl} \frac{1}{M^2} = \frac{1}{M^2} \delta_{j,k} \delta_{i,l}
\ee

which correspond to the totally local propagators\footnote{By a local propagator we mean here 
that it allows propagation only from a point to itself.}
\begin{equation}
<\phi(\tilde{p},x)\phi(\tilde{q},y)> \sim \delta(p+q)\delta(x-y)
\label{propphi}
\end{equation}

This corresponds, of course, to a situation where we have no
propagating degrees of freedom in the field theory. However, from
(\ref{propA}) it is clear that the $U$'s and $V$'s and the
phases that arise are the same as with the kinetic term.
Thus, for any given diagram, the question whether there is a phase
mixing between internal and external momenta, does not change when we
drop the kinetic term. We will use this observation later when we work
without the kinetic term to clarify the relation between non-planar
diagrams and IR physics.

\section{Origin of IR divergences}
\label{sec:divs}

The question we wish to answer in this section is -  where do the divergences 
of the non-commutative field theory come from, as viewed from the 
matrix (operator) formulation. 
We would like to understand the connection between the IR\footnote{By
  IR divergences we refer here only to divergences in some correlation
  functions, which diverge for vanishing  external momenta.} and UV
divergences,  and to reach a more intuitive explanation for the
connection between those IR effects and non-planar diagrams.

A finite dimensional matrix theory has no divergences, as it is a theory on a point. 
Infinities can only come from taking $N\rightarrow \infty$. Indeed, in a regular matrix 
theory the $N$ dependence is clear and one usually has a $1/N$ expansion of the physical 
quantities .The matrix theories arising from the non-commutative field theories have a 
kinetic term which complicates the $N$ dependence. 
Furthermore, the natural regularization and renormalization
procedures of the field theory throw away the leading $N$ behaviour. \newline
To get infra-red divergences in a correlation function one needs two
ingredients:
\begin{itemize}
\item An infinite volume
\item Long range interactions
\end{itemize}
We will see now how these conditions come about in the non-planar diagrams of 
a non-commutative field theory.

Let us first make the following observation: The size of the matrices
$N$, which plays
the role of a UV cut-off, as all momenta are in the range
$-\sqrt{N}< p < \sqrt{N}$, is also the configuration spaces area (at
fixed $\theta$), as

\begin{equation}
\mbox{Area}=\int d^2 x \sim Tr 1 = N.
\end{equation}
Hence the volume of space is proportional to $N$.
Therefore, any infra-red effect coming from the large volume of space time
are cut-off by $N$. This is one  of the reasons for the
IR$\leftrightarrow$UV relation, as
$N$ plays the role of both IR and UV cut-off's and there cannot be IR
divergences if we keep a finite UV cut-off. This, however does not explain why 
the IR divergences occur only when the theory has UV divergences (see below). 

\subsection{Long range interaction}
We would like to understand why certain non-planar diagrams 
are those which give
rise to the IR effects.
First, let us see how long range interactions are produced.
To gain insight into this question it turns out to be very useful to work not in momenta
space or configuration space but rather in the mixture of both. 
As explained in the previous sections, in this representation the star product
becomes simpler and more intuitive.
 Indeed, we have seen that the matrix description corresponds
to such a representation and the star product is then just matrix multiplication.
So we will work with fields that are functions of one configuration
and one momentum coordinate. To avoid annoying factors, we redefine the
matrix variables
\begin{equation}
A_{ij} \sim \phi(\tilde{i}-\tilde{j},i+j)
\end{equation}

To see the difference between an ordinary and a non-commutative theory let us consider 
the $\phi^4$ theory.
In an ordinary field theory the interaction vertex is local, $\int d^{2}x \phi^{4}(x)$.
This is written in a mixed configuration-momenta representation as
\begin{equation}
\int d^{4}pdx \prod_{i=1}^4 \phi(\tilde{p_i},x)\delta(\sum_{i=1}^{4}p_{i})
\end{equation}
where $\int d^4 p = \prod_{i=1}^4\int d p_i$.
On the other hand a non-commutative theory will have a non-local 
interaction of a special form.
For the quartic vertex the interaction takes the form (again, there
are some factors of 2 from the previous sections) 
\beq
\int d^{4}pdx \phi(\tilde{p_1},x)\phi(\tilde{p_2},x-p_1 -p_2)
\phi(\tilde{p_3},x-p_1-2p_2-p_3) \nonumber \\
\phi(\tilde{p_4},x-p_1-2p_2-2p_3-p_4)\delta(\sum_{i=1}^{4}p_{i}).
\eeq
The non-commutative theory has a non-local interaction whose 
non locality in one
direction depends on the momenta in the other direction, as was
noticed in \cite{BiSu}.
We will show that the long range interactions that are needed for an 
infra-red divergence 
arise from this non-local interaction, but only for certain non-planar
diagrams.

One can already guess how long range interaction may come about. We
have a non-local
interaction and a regular massive free propagator. Long range interactions require
that correlation functions are not exponentially suppressed for large distances 
(as they are in local massive theories). Such an effect can occur if 
the non-local interaction "draws" the interaction points close 
to each other, so while still having a massive propagator, the 
interaction is not exponentially suppressed. Given that, it seems clear that
the non-locality is proportional to the momenta flowing in the vertex, 
it is thus obvious that large non-locality is related to high momenta.

As a simple example let us consider  the two-point function in a non-commutative 
$\phi^4$ theory.
In an ordinary theory the first order (one loop) correction to 
the two point function is proportional to 
\begin{equation}
\int dx d^4p \; \phi(\tilde{p_1},x) \phi(\tilde{p_2},x)
<\phi(\tilde{p_3},x)\phi(\tilde{p_4},x)>_{0} \delta(\sum_{i=1}^4 p_i)
\end{equation}
Where $<>_{0}$ is the free propagator and we postpone the contraction
with the external $\phi$'s. Using the fact that the free propagator
has a momentum delta-function, we find

\be
\int dx dp_1 dp_2 \phi(\tilde{p_1},x) \phi(\tilde{-p_1},x)
<\phi(\tilde{p_2},x)\phi(\tilde{-p_2},x)>_{0} 
\ee
This term is, of course, the source for the UV divergences, as 
the fields interact at the same point.
We would like to see now what happens in the non-commutative theory. 
There are two possible corrections, depending on the contractions we
make in the vertex. If one contracts neighbouring fields, the diagram
is planar, while it is non-planar for contraction of non-neighbouring fields.
For the planar diagram one gets
\begin{eqnarray}
&\int& d^{4}pdx \phi(\tilde{p_1},x)
\phi(\tilde{p_2},x-p_1 -p_2)\nonumber\\
&\times& <\phi(\tilde{p_3},x-p_1-2p_2-p_3)
\phi(\tilde{p_4},x-p_1-2p_2-2p_3-p_4)>_{0}\delta(\sum_{i=1}^{4}p_{i})\nonumber
\end{eqnarray}
The free propagator is still proportional to a momentum delta-function
and the correlation function reduces to
\beqno
\int dx dp_1 dp_3 \phi(\tilde{p_1},x)\phi(-\tilde{p_1},x) 
 <\phi(\tilde{p_3},x+p_1-p_3)
\phi(-\tilde{p_3},x+p_1-p_3)>_{0} 
\eeqno
This is basically the same as the ordinary interaction coming from the local vertex.
On the other hand, the non-planar diagram gives 
\beqno
&\int& d^{4}pdx \phi(\tilde{p_1},x)\phi(\tilde{p_3},x-p_1-2p_2-p_3)\\
&\times& 
<\phi(\tilde{p_2},x-p_1 -p_2)\phi(\tilde{p_4},x-p_1-2p_2-2p_3-p_4)>_{0}
\delta(\sum_{i=1}^{4}p_{i})
\eeqno
Using momentum conservation we get now
\beqno
\int dp_1 dp_2 dx \phi(\tilde{p_1},x)\phi(-\tilde{p_1},x-2p_2) 
<\phi(\tilde{p_2},x-p_1 -p_2)\phi(-\tilde{p_2},x+p_1-p_2)>_{0}
\eeqno
We can see here that the non-planar diagram exhibits a different
behaviour. The non-planar contraction of two 'inner-legs' of 
the vertex, has caused the two 'outer-legs' to be in different points
in configuration space. These points are further away the larger the 
momentum
in the loop is. Thus, the long range forces are related to high momentum. 
We also see that the loop integral will not diverge unless $p_1=0$ as
the inner propagator is suppressed for any other value.

Let us look at another example, that of a cubic theory.
The interaction vertex is
\begin{eqnarray}
\int d^{3}p dx \phi(\tilde{p_1},x)\phi(\tilde{p_2},x-p_1 -p_2)
\phi(\tilde{p_3},x-p_1-2p_2-p_3)\delta(\sum_{i=1}^{3}p_{i}).
\end{eqnarray}

Consider, again, the one loop correction to the two point function. 
The planar diagram contribution is
\begin{eqnarray}
\int d^{3}p d^{3}q dx dy
\phi(\tilde{p_1},x)\phi(\tilde{q_1},y)<\phi(\tilde{p_2}, x-p_1 -p_2)
\phi(\tilde{q_{3}},y-q_2)>_{0}\nonumber\\
<\phi(\tilde{p_3},x-p_2)\phi(\tilde{q_2},y-q_1-q_2)>_{0}
\delta(\sum_{i=1}^{3}p_{i})\delta(\sum_{i=1}^{3}q_{i})
\end{eqnarray}
Using the various momentum delta-functions, we get
\beqno
\int dp_1 dp_2 dx dy \phi(\tilde{p_1},x)\phi(-\tilde{p_1},y)<\phi(\tilde{p_2},x-p_1 -p_2)
\phi(-\tilde{p_{2}},y-p_1-p_2)>_{0}\\
<\phi(-(\tilde{p_1} +\tilde{p_2}),x-p_2)\phi(\tilde{p_1} +\tilde{p_2},y-p_2)>_{0}
\eeqno
All propagators are ordinary free propagators and we see that they are 
evaluated at separation $(x-y)$ (although the points are shifted the propagator is a 
function of the difference of the points which remains
unchanged). Thus, for $x\neq y$ the  contribution from the propagators
is exponentially suppressed and there is no non-locality. In fact, 
this looks like an ordinary cubic one-loop correction

On the other hand, the non-planar diagram contribution is
\beqno
\int dp_1 dp_2 dx dy \phi(\tilde{p_1},x)\phi(-\tilde{p_1},y)<\phi(\tilde{p_2},x-p_1 -p_2)
\phi(-\tilde{p_{2}},y+p_1+p_2)>_{0}\\
<\phi(-(\tilde{p_1} +\tilde{p_2}),x-p_2)\phi(\tilde{p_1} +\tilde{p_2},y+p_2)>_{0}
\eeqno
Again, we see here a different behaviour. First, the inner propagators are evaluated
at configuration space distances $x-y-2p_1-2p_2$ and $x-y-2p_2$
respectively. Thus, even for $x\neq y$
the diagram is not always suppressed by the massive propagator. One can also see
why the divergences occur in the infra red, as the distances are equal
only when $p_1=0$. At any other case, at least one of the propagator
is suppressed. The same remarks as in the quartic coupling apply here.

To summarize, as the external legs in a non-planar diagram are located
further apart, the main contribution is due to higher and higher momenta.
Thus, integrating
over all configuration space, the large contributions come from large momenta
and it would look as if the infra-red effects are connected to UV 
physics.
Another point to notice is that, as the divergences of a diagram come from the high 
momentum contribution of the loop (as the IR divergences come from the summing over 
larger and larger separations), there will be an IR 
divergence only if this loop diverges and there are also UV divergences. 
This is because the only
source of divergences in perturbation theory is that of coincident points.
If there are no UV divergences there might still be non-local effects, but there
will not  be long range interactions, 
although there will be interaction on a larger scale
than the same local theory.
We have thus described the mechanism through which long range
interactions appear in non-commutative field theories.

\subsection{Why non-planar?}
We would like now to address the question of the relation between
non-planar diagrams and the IR physics from another point of
view. For that purpose, we return to the $\theta \rightarrow \infty$ limit
where there is no kinetic term in the action.

We claim that the essential point for our discussion - the appearance
of IR divergences - does not depend on the inclusion of the kinetic
term. There are several arguments in favour of this claim. As
explained in \cite{MiRaSe}, the IR divergences appear due to the phases
depending on the internal and external momenta. It was demonstrated in
section \ref{sec:feyn} that these phases indeed do not depend on the
inclusion of the kinetic term. Furthermore, if we choose the propagator to
 be totally local, the only non-local effects in any correlation
 function are those coming from the non-local interactions (rather
 than from any non-locality in the propagator), thus the non-local
 effects we are after are essentially the same, and even clearer, once
 the kinetic term is neglected. Finally, the propagator
 (\ref{fullphi}) of the full theory, with the kinetic term, is peaked
 around the values of the local (\ref{propphi}) and is exponentially
 suppressed away from these values.

Let us consider a general $n$-point correlation function in the theory
without the kinetic term.
\begin{equation}
< A_{i_1,j_1}A_{i_2,j_2}A_{i_3,j_3}\cdots A_{i_n,j_n}>
\label{corel}
\end{equation}

\begin{figure}[h!]
\begin{center}
\resizebox{0.6\textwidth}{!}{\includegraphics{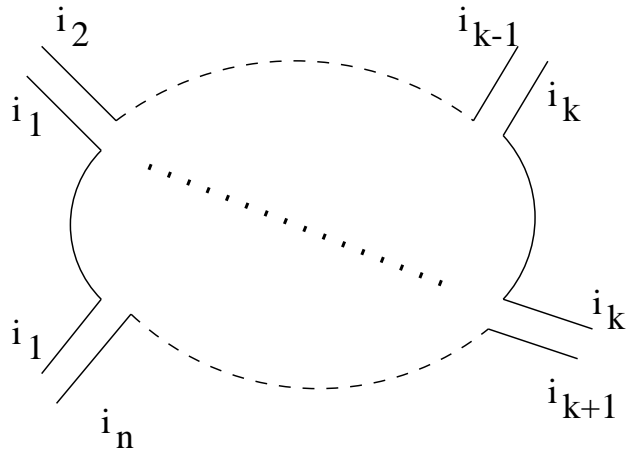}}
\end{center}
\caption{A general planar diagram.}
\label{fig:planar}
\end{figure}

The propagators (\ref{propA}) show us that we can use the two line
notation with a matrix index on each line. A general planar diagram,
depicted in figure (\ref{fig:planar}), contributes to the correlation function  
(\ref{corel}) only for the index structure
\begin{equation}
j_m=i_{m+1}\ \  mod\ \ n
\label{phplan}
\end{equation}
A non-planar diagram, on the other hand, may be of a more general
index structure.

We see here an important difference between planar and non-planar 
diagrams --  in a 
planar diagram there is only one group of indices that is connected in
a cyclic manner, while  a non-planar 
diagram can contribute even if the outside indices are separated 
into two or more groups. 
For example, the separation into two groups gives -
\beq
j_m&=&i_{m+1}\ \ \ \   m=1\cdots n_{1}-1,\ \ \  i_1=j_{n_1-1}.\nonumber \\
j_m&=&i_{m+1}\ \ \ \   m=n_{1}+1\cdots n-1,\ \ \  i_{n_1 +1}=j_{n}.
\label{phnonplan}
\eeq
Figure ({\ref{fig:nonpl}) is an example of such a diagram. 
We will now show that the structure of the phases is in one to one 
correspondence with the allowed index structure.

\begin{figure}[h!]
\begin{center}
\resizebox{0.6\textwidth}{!}{\includegraphics{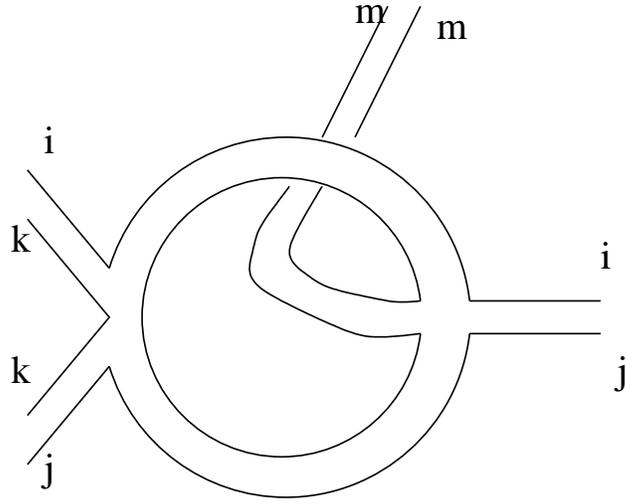}}
\end{center}
\caption{A non planar $\phi^4$ 4-points diagram.}
\label{fig:nonpl}
\end{figure}

Let us look at a contribution to the correlation function (\ref{corel}).
We will label the incoming momenta by $(m_l,n_l)$ and the inside momenta by
$(p_k,q_k)$. The contribution from any diagram is then
\be
\sum_{m_l,n_l,p_k,q_k}\left(\prod_{h}:U^{m_h}V^{n_h}:_{i_h,j_h}\right)
\tilde{f}(m_l,n_l.p_k,q_k)
\ee
Given the simple propagator and vertex, the function$\tilde{f}$ can only be made
 out of some delta functions of momenta and phases.
If the phases do not mix between internal and external momenta then
after summing over the internal momenta one gets only
\be
f(m_l,n_l) \equiv \sum_{p_k,q_k}\tilde{f}(m_l,n_l,p_k,q_K)= c N^{a}e^{i\alpha(m_l,n_l)}\delta_{(\sum_{l}m_l,0)}\delta_{(\sum_{l}n_l,0)}
\ee
where $\alpha(m_l,n_l)$ is some function of the external momenta and $c$ and
$a$ are constants.
If, however, there is some mixing between the internal and external
momenta then after summing over the internal momenta
one gets extra delta functions
\be
f(m_l,n_l)=c' N^{a'}e^{i\alpha'(m_l,n_l)}\delta_{(\sum_{l}m_l,0)}
\delta_{(\sum_{l}n_l,0)}
\cdots \delta_{(\sum'_{l}m_l,0)}\delta_{(\sum'_{l}n_l,0)}
\ee
where the primed sums indicate that the sum is not over all $l$.
We see that the structure of delta functions in $f(m_l,n_l)$ is determined
by the mixing (or non mixing) of the phases.

On the other hand, $f(m_l,n_l)$ is also determined by the allowed index 
structure, as one can just do an inverse transform by multiplying by
\be
\prod_{h'}:U^{m'_{h'}}V^{n'_{h'}}:_{i_{h'},j_{h'}},
\ee
and performing a trace.
This shows that there must be a one-to-one correspondence between the
index structure and the phase mixing.
As an example let us first look at a planar diagram.
The index structure is always  as in
equation (\ref{phplan}), so one finds
\be
f(m_l,n_l)\sim Tr \left(\prod_{h} :U^{m_{h}}V^{n_{h}}:
\right) \sim \delta_{(\sum_{l}m_l,0)}\delta_{(\sum_{l}n_l,0)}
\ee
On the other hand a non-planar diagram with an index structure as in equation
(\ref{phnonplan}) will give
\begin{eqnarray}
f(m_l,n_l)\sim Tr (\prod_{h=1}^{h_1} :U^{m_{h}}V^{n_{h}}: )
Tr (\prod_{h=h_{1}+1}^{n} :U^{m_h}V^{n_h}: ) \sim \nonumber\\
\delta_{(\sum^{}_{l}m_l,0)}\delta_{(\sum^{}_{l}n_l,0)}\delta_{(\sum^{h_1}_{l}m_l,0)}
\delta_{(\sum^{h_1}_{l}n_l,0)}
\end{eqnarray}

We have thus shown that phase mixing between internal and external 
momenta is equivalent to the grouping of the external indices. As a result,
the phase mixing can only occur for non-planar diagrams with a particular index structure.

Using the
relation of the matrix indices to the mixed momentum - configuration
space, we interpret this difference as follows.
In any $n$-point planar diagram, given the configuration space
coordinates of $(n-1)$ points, the coordinate of the last point is
fixed. IR effects are related to non-planar diagrams, because in some
non-planar diagrams some of the legs can be taken to be independently far from
the others (even if the propagator is totally local).  
The freedom in the index structure of non-planar diagram, given the
local form of the propagator, explains why the non-local effects discussed 
previously occur only for the non-planar diagrams.

As we claimed before that the classification of diagrams to planar and
non-planar in the momentum space representation of non-commutative
theories is the same as in the matrix formalism, it is clear now
why non-planar diagrams are connected to IR effects.

Let us consider the simplest example of this behaviour - the $\phi^4$ two point
function. Planar diagrams contribute only to (there is no summation on the indices)
\begin{equation}
<A_{ij}A_{ji}>
\label{p2}
\end{equation}
while non-planar can also contribute to 
\begin{equation}
<A_{ii}A_{jj}>.
\label{nonp2}
\end{equation}
Thus, even though we started with a totally localized propagator (i.e
proportional to $\delta(x-y)$)
the non-planar graphs may lead to a non-local correction, while the planar 
graphs may not. We see that this behaviour is totally encoded in the
index structure of the possible planar and  non-planar diagrams.
This is not to say that non-planar diagrams have to be of this form,
but that only non-planar diagrams that separate into two or more 
groupings of indices, are those that will exhibit the IR divergences.

Diagrams with the same index structure as planar ones will
have no phase mixing 
and will not exhibit IR divergences related to the values of the
outside momenta.

\subsection{The power of $N$}
One may still be interested in the degree of the IR divergences (or - how do the
non-planar diagram end up with the same power of $N$ as planar
ones). The degree of divergence
in the theory without the kinetic term, is simply the $N$
dependences. To consider that, let us look, once again, at the example
of the two point function in $\phi^{4}$ theory.
We will look at the contributions to the two-point function at zero momenta,
$<\tilde{A}(0,0) \tilde{A}(0,0)>$ where $\tilde{A}(0,0)=Tr A$.
The planar diagram contributes (at one loop) to terms of the form
\begin{equation}
\label{placor}
<A_{ij}A_{ji}>\sim \frac {1}{M^{2}}N \delta_{i,j}
\end{equation}
While the non-planar diagram contributes to terms of the form
\begin{equation}
\label{nonplacor}
<A_{ii}A_{jj}> \sim \frac{1}{M^{2}}
\end{equation}
where there is no summation in both cases. We see that the planar diagrams
 have a higher $N$ dependence than the non-planar ones.
However, there are $N$ planar terms that contribute to this two-point function , but 
$N^2$ non-planar terms, thus giving and overall identical $N$ dependence.

We can generalize this picture to any $n$-point function. Transforming
the planar diagram from the matrix space to momentum space we have a single
trace over $\Uop$'s and $\Vop$'s : 
\beq
<\tilde{\Aop}(p_1,q_1) ... \tilde{\Aop}(p_n,q_n)> &\sim & Tr (\Uop^{p_1}
  \Vop^{q_1} ... \Uop^{p_n} \Vop^{q_n}) ... \nonumber \\
  &= &N \cdot
  \delta_{p_1+...+p_n,0} \delta_{q_1+...+q_n,0} \times \mbox{phase}
  ...
\eeq
where the $\delta$ corresponds to the usual momentum conservation
$\delta$-function. A non-planar diagram, which might have a different grouping of the
external indices, will give
\beq
\lefteqn{<\tilde{\Aop}(p_1,q_1) ... \tilde{\Aop}(p_k,q_k)
\tilde{\Aop}(p_{k+1}q_{k+1}) ... \tilde{\Aop}(p_n,q_n)> \sim }
\nonumber \\
 & & Tr (\Uop^{p_1}  \Vop^{q_1} ... \Uop^{p_k} \Vop^{q_k}) 
\times  Tr (\Uop^{p_{k+1}}
  \Vop^{q_{k+1}} ... \Uop^{p_n} \Vop^{q_n}) ... =  \nonumber \\ 
 & & N^2 \cdot
  \delta_{p_1+...+p_k,0} \delta_{q_1+...+q_k,0}
  \delta_{p_{k+1}+...+p_n,0} \delta_{q_{k+1}+...+q_n,0} \times \mbox{phase}
  ... 
\eeq
The additional traces which  give additional factors of $N$ as well as
new kinematical restrictions are the hallmark of the possible IR
divergences. The corresponding factors of $N$ may cause these diagrams
to be of the same order in $N$ as the planar diagrams. 

The various factors of $N$ are also related to a fact briefly
mentioned before. As mentioned for example in \cite{MiRaSe}, the
$\theta \rightarrow \infty$ limit is dominated at generic external
momenta by the planar diagrams. This is obvious from equations
(\ref{placor}) and (\ref{nonplacor}). However, at special values of
the external momenta, this dominance may be compensated by the
additional traces and we may encounter non-planar diagrams which
contribute leading (and even diverging) terms.
 
\appendix
\section{Appendix}
As a demonstration of the origin of the phases in the operators
formalism, we give a simple example - consider the one loop, 
two point function in non-commutative $\phi^3$. The expectation value
to be computed is 
\be
<\Aop_{ab} \Aop_{cd} Tr \Aop^3 Tr \Aop^3> = <\Aop_{ab} \Aop_{cd}
\Aop_{a_1b_1} \Aop_{b_1c_1} \Aop_{c_1a_1} \Aop_{a_2b_2} \Aop_{b_2c_2}
\Aop_{c_2a_2} >
\ee
There are two distinct Feynman graphs corresponding to this
expression, a planar graph and a non planar one (see figure
\ref{fig:graphs}). 

\begin{figure}[h!]
\begin{center}
\resizebox{0.6\textwidth}{!}{\includegraphics{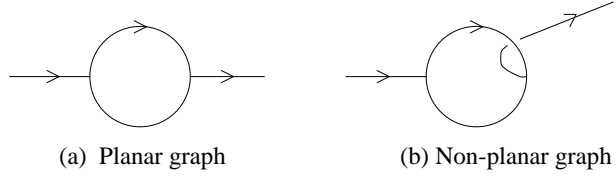}}
\end{center}
\caption{one-loop Feynman graphs.}
\label{fig:graphs}
\end{figure}
The planar graph is given by the contraction

\beqno
\lefteqn{ <\Aop_{ab} \Aop_{a_1b_1}><\Aop_{b_1c_1}
\Aop_{a_2b_2}><\Aop_{c_1a_1}\Aop_{c_2a_2}><\Aop_{b_2c_2} \Aop_{cd}>=} \\
 & \sum_{m_in_i} & 
 :\Uop^{m_1} \Vop^{n_1}:_{ab} :\Uop^{-m_1}\Vop^{-n_1 }:_{a_1b_1}  :\Uop^{m_2}
 \Vop^{n_2}:_{b_1c_1} :\Uop^{-m_2}\Vop^{-n_2 }:_{a_2b_2} \\ 
 & &:\Uop^{m_3} \Vop^{n_3}:_{c_1a_1} :\Uop^{-m_3}\Vop^{-n_3
   }:_{c_2a_2}  :\Uop^{m_4}
 \Vop^{n_4}:_{b_2c_2} :\Uop^{-m_4}\Vop^{-n_4 }:_{cd} ... = \\
 & \sum_{m_in_i} & 
:\Uop^{m_1} \Vop^{n_1}:_{ab} :\Uop^{-m_4} \Vop^{-n_4}:_{cd} Tr
\left(:\Uop^{-m_1} \Vop^{-n_1}::\Uop^{m_2} \Vop^{n_2}::\Uop^{m_3}
  \Vop^{n_3}: \right) \\
 & & Tr \left(:\Uop^{-m_2} \Vop^{-n_2}::\Uop^{m_4}
   \Vop^{n_4}::\Uop^{-m_3} \Vop^{-n_3}: \right) ... = \\ 
 & \sum_{m_in_i} & 
:\Uop^{m_1} \Vop^{n_1}:_{ab} :\Uop^{-m_4} \Vop^{-n_4}:_{cd} \times 
 \exp(i \theta ( m_1n_1/2 + m_2n_2 + m_3n_3 + \\
 & &  m_4n_4/2 - m_2n_1 -
   m_3n_1 - m_4n_2 - m_3n_4 +  2m_3n_2)) \times \\
 & & \delta(m_1-m_2-m_3) \delta(m_4-m_2-m_3)
\delta(n_1-n_2-n_3) \delta (n_4-n_2-n_3) ... 
\eeqno  

where $...$ represent further terms that do not contribute to the
phase. It can be easily seen that the phase in the last line adds up
to zero and one is left only with the phase coming from the external
legs.

The non-planar graph is given by 

\beqno
\lefteqn{ <\Aop_{ab} \Aop_{a_1b_1}><\Aop_{b_1c_1}
\Aop_{a_2b_2}><\Aop_{c_1a_1}\Aop_{b_2c_2}><\Aop_{c_2a_2} \Aop_{cd}>=} \\
 & \sum_{m_in_i} & 
 :\Uop^{m_1} \Vop^{n_1}:_{ab} :\Uop^{-m_1}\Vop^{-n_1 }:_{a_1b_1}  :\Uop^{m_2}
 \Vop^{n_2}:_{b_1c_1} :\Uop^{-m_2}\Vop^{-n_2 }:_{a_2b_2} \\ 
 & &:\Uop^{m_3} \Vop^{n_3}:_{c_1a_1} :\Uop^{-m_3}\Vop^{-n_3
   }:_{b_2c_2}  :\Uop^{m_4}
 \Vop^{n_4}:_{c_2a_2} :\Uop^{-m_4}\Vop^{-n_4 }:_{cd} ... = \\
 & \sum_{m_in_i} & 
:\Uop^{m_1} \Vop^{n_1}:_{ab} :\Uop^{-m_4} \Vop^{-n_4}:_{cd} Tr
\left(:\Uop^{-m_1} \Vop^{-n_1}::\Uop^{m_2} \Vop^{n_2}::\Uop^{m_3}
  \Vop^{n_3}: \right) \\
 & & Tr \left(:\Uop^{-m_2} \Vop^{-n_2}::\Uop^{-m_3}
   \Vop^{-n_3}::\Uop^{m_4} \Vop^{n_4}: \right) ... = \\ 
 & \sum_{m_in_i} & 
:\Uop^{m_1} \Vop^{n_1}:_{ab} :\Uop^{-m_4} \Vop^{-n_4}:_{cd} \times 
  \exp(i \theta ( m_1n_1/2 + m_2n_2 + m_3n_3 \\
 & & +  m_4n_4/2 - m_2n_1 -
   m_3n_1 - m_4n_3 - m_4n_2 +  2m_3n_2)) \times \\
 & & \delta(m_1-m_2-m_3) \delta(m_4-m_2-m_3)
\delta(n_1-n_2-n_3) \delta (n_4-n_2-n_3) ... = \\
  & \sum_{m_in_i} & 
:\Uop^{m_1} \Vop^{n_1}:_{ab} :\Uop^{-m_4} \Vop^{-n_4}:_{cd} \exp{(i
  \theta (- m_2 n_3 + m_3n_2))} ... 
\eeqno  

giving the same phase as the corresponding diagram in
momentum space.

\end{document}